# 2-D Analysis of Enhancement of Analytes Adsorption Due to Flow Stirring by Electrothermal Force in The Microcantilever Sensor

Ming-Chih Wu[1], Jeng-Shian Chang[2], Chih-Kai Yang[1]
**Institute of Applied Mechanics, National Taiwan University**
jschang@spring.iam.ntu.edu.tw

[1] Graduate student
[2] Professor

**ABSTRACT**

Ac electrokinetic flows are commonly used for manipulating micron-scale particles in a biosensor system. At the solid-liquid state there are two kinds of processes in the reaction between analytes and ligands: the mass transport process and the chemical reaction process. The mass transport process is related to convection and diffusion. Total or partial limit of mass transport would retard the diffusion from the bulk fluid to the interface of reaction. This effect decreases the possibility of adsorption of analyte and ligand because the chemical reaction is faster than the diffusion. In order to solve this problem, we apply an ac electric field to induce a vortex field by the electrothermal effect, which helps in increasing the rate of diffusion. By using the finite element analysis software, COMSOL Multiphysics $^{TM}$ (COMSOL Ltd., Stockholm, Sweden), we optimized several parameters of the microelectrode structures with a simplified 2-D model.

## 1. INTRODUCTION

Sensors are extensively used in our daily lives. In recent years biosensors have become a hot research field for its high sensitivity and real-time detecting ability. The three most common devices in detecting tiny molecules are the microcantilever biosensor, the SPR (Surface Plasmon Resonance) sensor, and the QCM (Quartz Crystal Microbalance) sensor. The microcantilever sensor can satisfy the demand for microminiaturization, but this device takes a lot of time during experiments. One of the reasons is that the binding of the analytes and the ligands is restrained by mass-transport limit, which leads to diffusion layers. In many cases, it's desirable to manipulate particle movement in a more efficient way. Ac electrokinetics refers to induced particle and fluid motion resulting from externally applied ac electric fields. This paper discusses how the ac electrokinetic force affects in the reacting rate of the analyte and the ligand, which have the size of sub-micrometer.

Microelectrode structures are commonly used in ac electrokinetics to generate the high strength ac electric fields, which is required in moving suspended particles in liquid [1]. A wide range of sizes of particles have been dielectrophoretically manipulated in this manner, from cells (~10 μm) and bacteria down to viruses (~100 nm) and protein molecules [2, 3]. Ac electrokinetics can be classified as three kinds of force: dielectrophoresis , electrothermal force , and electro-osmosis [4]. While a nonuniform ac electric field can move suspended particles using dielectrophoretic forces, it can also move the fluid through the electrothermal effect or ac electro-osmosis [5, 6, 7]. Ac electro-osmosis is only influential at the frequency below 10 KHz. Also, it is likely to hydrolyze at lower frequency [8]. Dielectrophoresis doesn't obviously affect the motion of

  

the particle in the scale of sub-micrometer. In contrast, the electrothermal effect operating in higher frequency is dominant in the bulk fluid [9]. Hence only the electrothermal force is discussed in this paper.

By numerical simulation, we apply an ac electric field to induce Joule heating. The induced electrothermal force will construct a vortex field which can reduce the thickness of the diffusion layer. We expect the reduction of the thickness of the diffusion layer can accelerate the reacting rate to shorten the experimental time. However, this method is only suitable for the molecules which are restrained by mass transport limit. Simulations of the surface concentration of the complex are performed in this paper.

## 2. Theory

At the solid-liquid interface, the reaction has two step events [10].

1. Mass transport process: the analyte is transferred out of the bulk solution towards the reacting surface.

   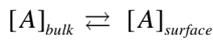

2. Chemical reaction process: the binding of the analyte to the ligand takes place.

   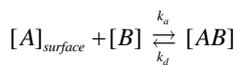

where $[A]_{bulk}$ is the analyte in the bulk, $[A]_{surface}$ is the analyte at the gold surface with ligand bound to the dextran matrix, $[B]$ is the ligand, $[AB]$ is the analyte-ligand complex, $k_a$ is the association rate constant, and $k_d$ is the dissociation rate constant.

When the analyte takes a longer time to convect and diffuse to the reacting surface than chemical reaction, the whole reaction is restrained by mass transport limit. It usually occurs in a lower flow speed, smaller diffusion coefficient, and higher association rate constant (greater than $10^6 M^{-1}s^{-1}$) situation.

Accordingly, we apply an ac electric field in a micro fluidic channel to produce vortex field in order to reduce the thickness of the diffusion layer. Hence we can shorten the experimental time by accelerating the reacting rate.

### 2.1 Electrothermal force

Electrothermal force is induced by Joule heating which arises from the inhomogeneties in the electric field. Accordingly, it transforms electrical energy into heat energy to increase the temperature gradients. We call it the thermal effect of current or effect of Joule heating [4].

The gradient of temperature $T$ in the liquid causes inhomogeneities in the permittivity $\varepsilon$ and conductivity $\sigma$ of the medium, which give rise to forces causing fluid motion.

The body force $\vec{F_E}$ can be given by [4]:

$$\vec{F_E} = -\frac{1}{2}\left[\left(\frac{\nabla\sigma}{\sigma} - \frac{\nabla\varepsilon}{\varepsilon}\right)\cdot\vec{E}\frac{\varepsilon\vec{E}}{1+(\omega\tau)^2} + \frac{1}{2}|\vec{E}|^2\nabla\varepsilon\right] \quad (2.1)$$

$$= -\frac{\varepsilon}{2}\left[0.024\nabla T \cdot \vec{E}\frac{\vec{E}}{1+(\omega\tau)^2} + \frac{|\vec{E}|^2}{2}(-0.004)\nabla T\right] \text{ (for water)}$$

where $\tau = \varepsilon/\sigma$ is the charge relaxation time, and $\omega$ is angular frequency of the electric field $\vec{E}$.

The local variations in temperature change the gradients of conductivity and permittivity:

$$\nabla\varepsilon = (\partial\varepsilon/\partial T)\nabla T \quad (2.2)$$

$$\nabla\sigma = (\partial\sigma/\partial T)\nabla T \quad (2.3)$$

For water, $\frac{1}{\varepsilon}\left(\frac{\partial\varepsilon}{\partial T}\right) = -0.4\%$, $\frac{1}{\sigma}\left(\frac{\partial\sigma}{\partial T}\right) = 2\%$

per degree Kelvin, (Lide ,1994).

The force induced by a permittivity gradient is the dielectric force. The force induced by a conductivity gradient is the Coulomb force. If $\omega \ll \sigma/\varepsilon$ the force is dominated by the Coulomb force. If $\omega \gg \sigma/\varepsilon$ the force is dominated by the dielectric force.

### 2.2 The electric field

Because the electrothermal force is a time-averaged entity, it is sufficient to solve the static electric field that correspond to the root mean square (rms) value of the ac field. The electrostatics problem is solved with Laplace's equation [11].

$$\nabla^2 \Phi = 0 \quad (2.4)$$
$$\vec{E} = -\nabla \Phi \quad (2.5)$$

where $\Phi$ is the electrical potential.

## 2.3 The temperature field

The generation of this amount of Joule heating in a very small volume could give rise to a temperature increase in the fluid. In order to estimate the temperature rise for a given electrode array, the energy balance equation must be solved [12].

$$\rho c_p \frac{\partial T}{\partial t} + \rho c_p \vec{V} \cdot \nabla T = k \nabla^2 T + \sigma \left|\vec{E}\right|^2 \quad (2.6)$$

where $\rho$, $c_p$, and $k$ are the density of the fluid, specific heat, and thermal conductivity of the fluid, respectively. $\sigma \left|\vec{E}\right|^2$ is Joule heating.

## 2.4 The flow field

We assume that the density $\rho$ and viscosity $\eta$ of the modeled fluid are constant, which not be affected by temperature and concentration.

The governing equations of continuity and three-dimensional momentum can be expressed as follows:

$$\nabla \cdot \vec{V} = 0 \quad (2.7)$$
$$\rho \frac{\partial u}{\partial t} + \rho \vec{V} \cdot \nabla u - \eta \nabla^2 u + \frac{\partial p}{\partial x} = F_{E,x} \quad (2.8)$$
$$\rho \frac{\partial v}{\partial t} + \rho \vec{V} \cdot \nabla v - \eta \nabla^2 v + \frac{\partial p}{\partial y} = F_{E,y} \quad (2.9)$$
$$\rho \frac{\partial w}{\partial t} + \rho \vec{V} \cdot \nabla w - \eta \nabla^2 w + \frac{\partial p}{\partial z} = F_{E,z} \quad (2.10)$$

where $u$, $v$, $w$ are the velocity components in $x$, $y$, $z$ direction, respectively, $\eta$ is dynamic viscosity of the fluid, $\rho$ is density of the fluid, and $p$ is pressure.

## 2.5 The concentration field

Transport of protein to and from the surface is caused by fluid flow and by diffusion.

In order to solve the problem, we apply Fick's second law:

$$\frac{\partial [A]}{\partial t} + u\frac{\partial [A]}{\partial x} + v\frac{\partial [A]}{\partial y} + w\frac{\partial [A]}{\partial z} = D(\frac{\partial^2 [A]}{\partial x^2} + \frac{\partial^2 [A]}{\partial y^2} + \frac{\partial^2 [A]}{\partial z^2}) \quad (2.11)$$

where $[A]$ is the concentration of analyte, and D is the diffusion coefficient.

The relation between concentration flux at the surface and the rate of reaction is given by

$$-D\left(\frac{\partial [A]}{\partial z}\right)_{surface} = k_a [A]_{surface} \left\{[B]_0 - [AB]\right\} - k_d [AB] \quad (2.12)$$

where $[A]_{surface}$ is the concentration of the analyte at the reacting surface with ligand bound to the dextran matrix, $[B]_0$ is the unit thickness concentration (surface concentration) of the ligand, $[AB]$ is the unit thickness concentration (surface concentration) of the analyte-ligand complex, $k_a$ is the association rate constant, and $k_d$ is the dissociation rate constant.

## 2.6 The reaction surface

The reaction between immobilized ligand and analyte can be assumed to follow the first order Langmuir adsorption model [13, 14]. During the association phase, the complex $[AB]$ increases as a function of time according to:

$$\frac{\partial [AB]}{\partial t} = K_a [A]_{surface} \left\{[B]_0 - [AB]\right\} - K_d [AB] \quad (2.13)$$

## 3. Set-up of the simulation.

The simplified 2-D model is shown in Fig.3.1. Note that the dimension of the microcantilever beam is $40 \mu m \times 4 \mu m$.

In order to simplify the calculation, we use a 2-D model to optimize the design. Table 3.1 shows that the general parameters in our simulation. Fig 3.2 shows the mesh we used for the analysis. The degree of freedom and the number of elements are about 127000 and 13000, respectively. The convergence tests are performed well, so we consider that the results are convincible.

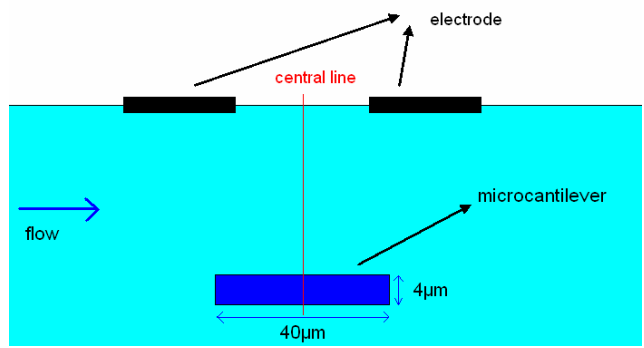

**Fig. 3.1 2-D model**

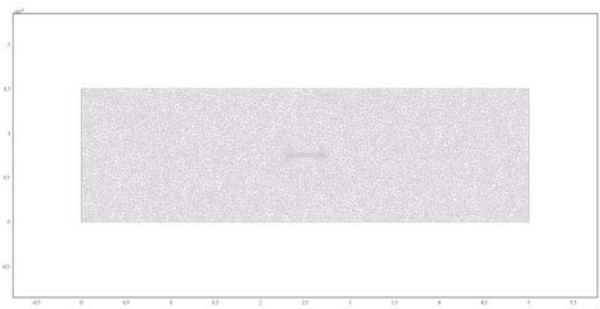

**Fig.3.2 The mesh of the 2-D model**

**Table 3.1 Parameters of the simulation**

| | |
|---|---|
| Flow speed (m/s) | $10^{-4}$ |
| Temperature of electrodes (K) | 300 |
| Coefficient of diffusion (m$^2$/s) | $10^{-10}$ |
| $k_a$ (M$^{-1}$s$^{-1}$) | 2600 |
| $k_d$ (s$^{-1}$) | 0.01 |
| Concentration of analyte [A]$_{surface}$ (M) | $10^{-5}$ |
| Concentration of ligand [B]$_0$ (mole/m$^2$) | $3\times10^{-8}$ |
| Fluid Density ρ (Kg/m$^3$) | $10^3$ |
| Relative permittivity ε$_r$ | 80.2 |
| Dynamic viscosity η (Pa・s) | $10^{-3}$ |
| Electrical conductivity σ (S・m$^{-1}$) | $5.75\times10^{-2}$ |

## 4. Results

### 4.1 Design of the width of the electrode

The width of the electrode will affect the amplitude and the distribution of the electric field. Furthermore the electrothermal force and the distribution of the flow field will be influenced. The width of the microcantilever is 40 ($\mu m$). We set the width of electrode as 60 ($\mu m$), 65 ($\mu m$), and 70 ($\mu m$) for simulation. The centerline lies on the center of the gap between two electrodes. We put the reacting surface, i.e. the microcantilever, at the central position to see if there is any influence in changing the width of the electrodes. Fig 4.1 shows that the width of electrodes does not affect the concentration of the complex conspicuous. We regard the width of 60 ($\mu m$) as the optimal parameter.

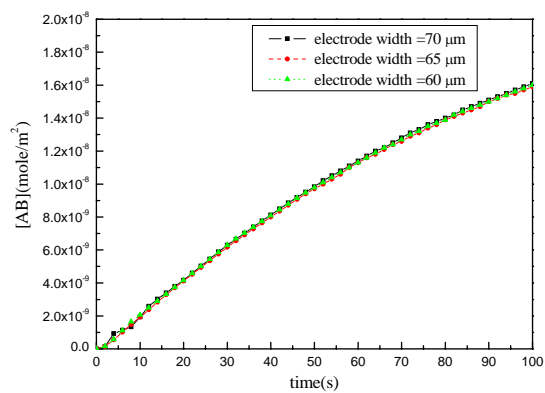

**Fig. 4.1 concentration of the complex (in different**

widths) as a function of time

### 4.2 Design of the gap between electrodes

The gap between electrodes would affect the amplitude and the distribution of the electric field, which may change the amplitude of the electrothermal force and the distribution of the flow field. We consider the distance of the gap as $10\,(\mu m)$, $15\,(\mu m)$, $20\,(\mu m)$ in the simulation. The reacting surface, i.e. the microcantilever, is also put at the central position between two electrodes. Fig 4.2 shows that the distance of the gap between electrodes does not affect the concentration of the complex conspicuously. We regard the gap of $15\,(\mu m)$ as the optimal parameter in this analysis.

### 4.3 Design of operating frequency

It's agreed that the electro-osmosis is only influential at the frequency below 10 KHz. So we simulate the concentration of the complex at the operating frequency as $10^2, 10^3, 10^4, 10^5, 10^6$ KHz with 25 $V_{rms}$. The reacting surface, i.e. the microcantilever, is also put at the central position between two electrodes. Fig 4.3 shows that it takes the least time to reach steady state when the frequency is $10^2$ or $10^3$ KHz. We set the frequency of $10^2$ KHz as the optimal operating frequency in the analysis.

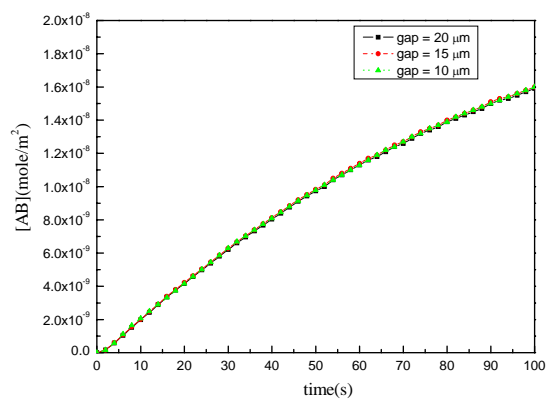

Fig. 4.2 concentration of the complex (in different gaps, which are the distances between electrodes) as a function of time

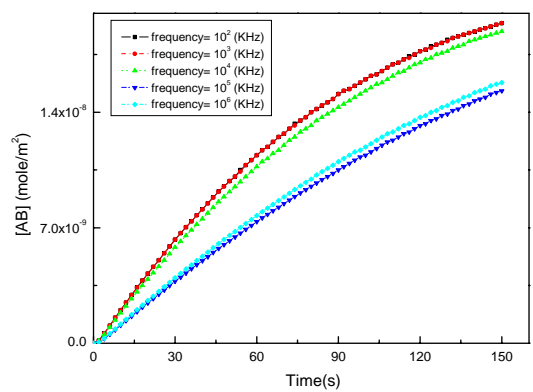

Fig. 4.3 concentration of the complex (in different operating frequencies) as a function of time

### 4.4 Effect of the amplitude of voltage

Table 4.1 shows the optimal parameters in our design. Note that the width of the electrode and the distance of gap are 60 μm and 15 μm, respectively. The reacting surface, i.e. the microcantilever, is also put at the central position between two electrodes. Then we apply different voltages: 5 V, 10 V, 15 V, 20 V, and 25 V to evaluate which is the most efficient. The result is shown in Fig. 4.5. Table 4.2 shows the numerical results of the simulation. When the operating voltage is 25 V, the fastest velocity in the flow field is $3.43 \times 10^{-2}$ m/s. Also its time to reach steady state is 426 seconds, which is less than the time in 0 V by 304 seconds. The factor of the efficiency (the ratio of the steady time in 0 V to 25 V) is about 1.71 when we applied an electric field of 25 $V_{rms}$ peak-to-peak. Fig. 4.6-9 show the post processing results for this model with the optimal parameters at the time of 450s.

| Width of electrode | Gap between electrodes | Frequency |
|---|---|---|
| 60 μm | 15 μm | $10^2$ KHz |

**Table 4.1 Optimal parameters in applying ac field**

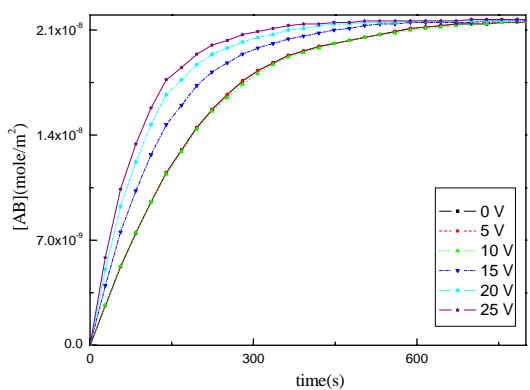

Fig. 4.5 concentration of the complex (in different voltages) as a function of time

Table 4.2 2-D simulation for different voltages

| Voltage (V) | Difference in temperature (K) | Max. downward velocity in $z$-axis (m/s) | Max. velocity (m/s) | Time to reach steady state (s) |
|---|---|---|---|---|
| 0 | 0 | $3.64 \times 10^{-5}$ | $1.53 \times 10^{-4}$ | 730 |
| 5 | 0.304 | $3.64 \times 10^{-5}$ | $1.56 \times 10^{-4}$ | 728 |
| 10 | 1.223 | $1.99 \times 10^{-4}$ | $8.05 \times 10^{-4}$ | 740 |
| 15 | 2.802 | $1.03 \times 10^{-3}$ | $4.09 \times 10^{-3}$ | 582 |
| 20 | 5.192 | $3.42 \times 10^{-3}$ | $1.35 \times 10^{-2}$ | 482 |
| 25 | 8.430 | $8.90 \times 10^{-3}$ | $3.43 \times 10^{-2}$ | 426 |

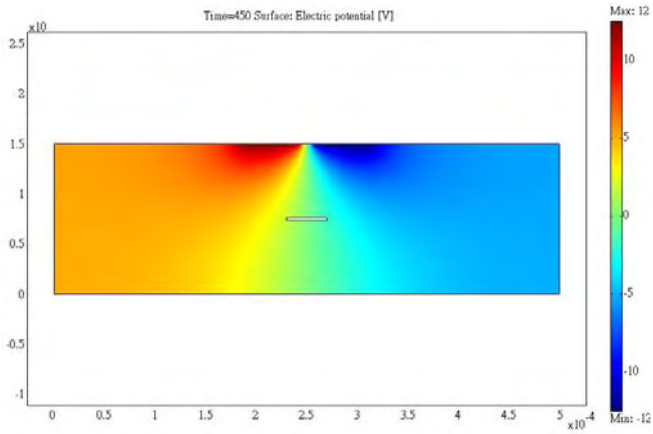

**Fig. 4.6** The distribution of electrical potential at the voltage of 25 V.

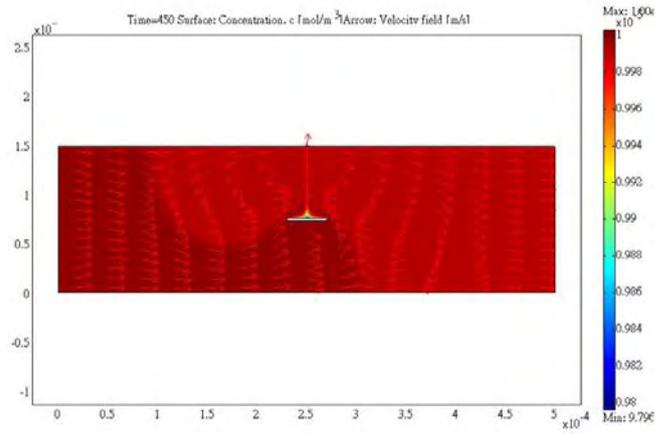

**Fig. 4.8** The distribution of $[A]_{surface}$ and the flow field in the channel at the voltage of 25 V.

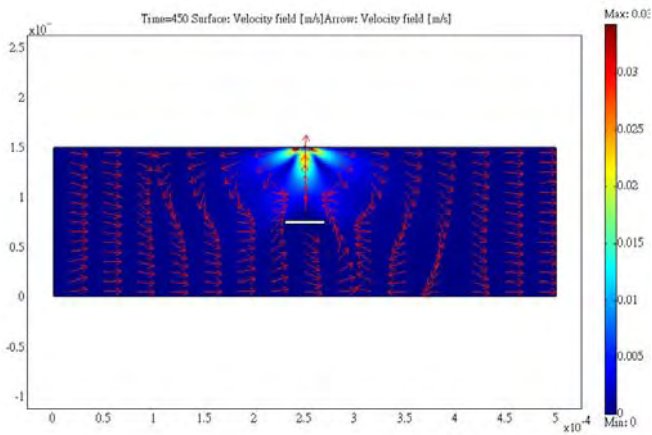

**Fig. 4.7** The distribution of flow field at the voltage of 25 V.

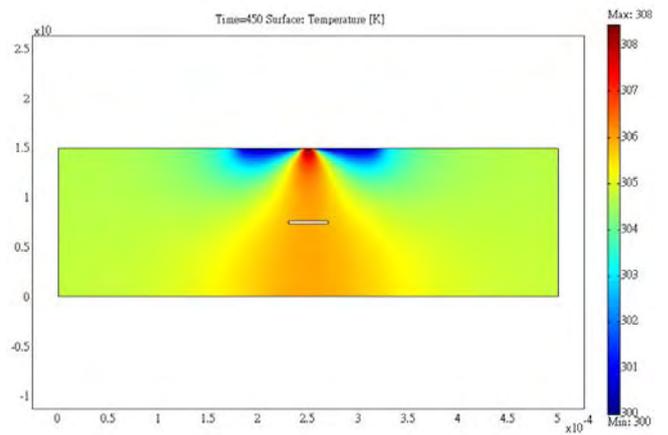

**Fig. 4.9** The distribution of temperature field at the voltage of 25 V.

## 5. CONCLUSION

In this paper, the simulation is performed by the finite element analysis software, COMSOL Multiphysics ™ (COMSOL Ltd., Stockholm, Sweden). We have already optimized several parameters of the microelectrode structures. The optimal widths and gap of the electrodes are 60 μm and 15 μm, respectively. Furthermore, the optimal operating frequency and voltage are 100 KHz and 25 V$_{rms}$ peak-to peak, respectively. These results can be a reference for advanced experiments or simulations.

It is successful in accelerating the reacting rate of the molecule which is limited by mass transport. The factor of the efficiency is about 1.71 when the operating voltage is 25 $V_{rms}$ peak-to-peak. In addition, the surface concentration of the complex on the microcantilever has been simulated. The future work is to simulate this project with a 3-D model. We expect that the results of 3-D simulation will be more accurate than 2-D. The 3-D simulation is being performed as we planned.

## 6. ACKNOWLEDGEMENTS

This research was supported by the National Science Council in Taiwan through NSC 94-2120-M-002-014.